\documentclass[useAMS,usenatbib]{mnras}
\usepackage{graphicx}
\usepackage{amsmath}
\usepackage{amssymb}
\usepackage{color}
\usepackage{multirow}

\def \chisq  {\ifmmode  \chi^2   \else  $\chi^2$  \fi}  
\def \spose#1{\hbox  to 0pt{#1\hss}} 
\def \lta{\mathrel{\spose{\lower 3pt\hbox{$\sim$}}\raise  2.0pt\hbox{$<$}}}
\def \gta{\mathrel{\spose{\lower  3pt\hbox{$\sim$}}\raise 2.0pt\hbox{$>$}}}

\def \kms {\ifmmode  \,\rm km\,s^{-1} \else $\,\rm km\,s^{-1}  $ \fi }
\def \kpc {\ifmmode  {\rm~kpc}  \else ${\rm~kpc}$\fi}  
\def \pc {\ifmmode  {\rm~pc}  \else ${\rm~pc}$ \fi  }  
\def \Gyr {\ifmmode  {\rm~Gyr}  \else ${\rm~Gyr}$\fi}
\def \Msun {\ifmmode {\rm M}_{\odot} \else ${\rm M}_{\odot}$ \fi} 
\def \Lsun {\ifmmode L_{\odot} \else $L_{\odot}$ \fi} 
\def \Rsun {\ifmmode R_{\odot} \else $R_{\odot}$ \fi} 
\def \Msunpyr {\ifmmode M_{\odot}{\rm~yr}^{-1} \else $M_{\odot}{\rm~yr}^{-1}$ \fi} 
\def \hMsun {\ifmmode h^{-1}\,\rm M_{\odot} \else $h^{-1}\,\rm M_{\odot}$ \fi}
\def \hMpc {\ifmmode  {h^{-1}\rm Mpc}  \else ${h^{-1}\rm Mpc}$ \fi  }  
\def \Mpch {\ifmmode  {h^{-1}\rm Mpc}  \else ${h^{-1}\rm Mpc}$ \fi  }

\def \LCDM {\ifmmode \Lambda{\rm CDM} \else $\Lambda{\rm CDM}$ \fi}
\def \sig8 {\ifmmode \sigma_8 \else $\sigma_8$ \fi} 
\def \OmegaM {\ifmmode \Omega_{\rm M} \else $\Omega_{\rm M}$ \fi} 
\def \OmegaL {\ifmmode \Omega_{\rm \Lambda} \else $\Omega_{\rm \Lambda}$\fi} 
\def \Deltavir {\ifmmode \Delta_{\rm vir} \else $\Delta_{\rm vir}$ \fi}
\def \rhocrit {\ifmmode \rho_{\rm crit} \else $\rho_{\rm crit}$ \fi}
\def \rhou {\ifmmode \rho_{\rm u} \else $\rho_{\rm u}$ \fi}
\def \zc {\ifmmode z_{\rm c} \else $z_{\rm c}$ \fi}

\def\LCDM{$\Lambda$CDM }
\def \LCDM {\ifmmode \Lambda{\rm CDM} \else $\Lambda{\rm CDM}$ \fi}
\def \sig8 {\ifmmode \sigma_8 \else $\sigma_8$ \fi} 
\def \Omegam {\ifmmode \Omega_{\rm m} \else $\Omega_{\rm m}$ \fi} 
\def \Omegab {\ifmmode \Omega_{\rm b} \else $\Omega_{\rm b}$ \fi} 
\def \Omegar {\ifmmode \Omega_{\rm r} \else $\Omega_{\rm r}$ \fi} 
\def \fbar {\ifmmode f_{\rm bar} \else $f_{\rm bar}$ \fi} 
\def \OmegaL {\ifmmode \Omega_{\rm \Lambda} \else $\Omega_{\rm \Lambda}$\fi} 
\def \Deltavir {\ifmmode \Delta_{\rm vir} \else $\Delta_{\rm vir}$ \fi}
\def \rhocrit {\ifmmode \rho_{\rm crit} \else $\rho_{\rm crit}$ \fi}

\def \rhos {\ifmmode \rho_{\rm s} \else $\rho_{\rm s}$ \fi} 
\def \rs {\ifmmode r_{\rm s} \else $r_{\rm s}$ \fi} 
\def \cvir {\ifmmode c_{\rm vir} \else $c_{\rm vir}$ \fi} 
\def \Rvir {\ifmmode r_{\rm vir} \else $R_{\rm vir}$ \fi}
\def \Vvir {\ifmmode V_{\rm  vir} \else  $V_{\rm vir}$  \fi} 
\def \Mvir {\ifmmode M_{\rm  vir} \else $M_{\rm  vir}$ \fi}  
\def \Nvir {\ifmmode N_{\rm  vir} \else $N_{\rm  vir}$ \fi}  
\def \Jvir {\ifmmode J_{\rm vir} \else $J_{\rm vir}$ \fi} 
\def \Evir {\ifmmode E_{\rm vir} \else $E_{\rm vir}$ \fi} 
\def \vvir {\ifmmode v_{\rm vir} \else $v_{\rm vir}$ \fi} 
\def \lam {\ifmmode \lambda  \else $\lambda$ \fi} 
\def \lamp {\ifmmode \lambda^{\prime} \else $\lambda^{\prime}$  \fi} 
\def \Vmax {\ifmmode V_{\rm  max} \else  $V_{\rm max}$  \fi} 
\def \Mdm {\ifmmode M_{\rm  dm} \else $M_{\rm  dm}$ \fi}

\def \Mgas {\ifmmode M_{\rm gas} \else $M_{\rm gas}$ \fi} 
\def \Mcg {\ifmmode M_{\rm cg} \else $M_{\rm cg}$\fi} 
\def \Mhg {\ifmmode M_{\rm hg} \else $M_{\rm hg}$ \fi} 
\def \Mdisc {\ifmmode M_{\rm disc} \else $M_{\rm disc}$ \fi} 
\def \Md {\ifmmode M_{\rm d} \else $M_{\rm d}$ \fi} 
\def \Mda {\ifmmode M_{\rm d,0\%} \else $M_{\rm d,0\%}$ \fi} 
\def \Mdb {\ifmmode M_{\rm d,20\%} \else $M_{\rm d,20\%}$ \fi} 
\def \Mdc {\ifmmode M_{\rm d,40\%} \else $M_{\rm d,40\%}$ \fi} 
\def \md {\ifmmode m_{\rm d} \else $m_{\rm d}$ \fi} 
\def \Mb {\ifmmode M_{\rm b} \else $M_{\rm b}$ \fi} 
\def \Mbh {\ifmmode M_{\rm b,pri} \else $M_{\rm b,pri}$ \fi} 
\def \Mbs {\ifmmode M_{\rm b,sat} \else $M_{\rm b,sat}$ \fi} 
\def \zo {\ifmmode z_{0} \else $z_{0}$ \fi} 
\def \rd {\ifmmode r_{\rm d} \else $r_{\rm d}$ \fi}
\def \rg {\ifmmode r_{\rm g} \else $r_{\rm g}$ \fi}
\def \rb {\ifmmode r_{\rm b} \else $r_{\rm b}$\fi}
\def \rs {\ifmmode r_{\rm s} \else $r_{\rm s}$\fi}
\def \rc {\ifmmode r_{\rm c} \else $r_{\rm c}$\fi}
\def \rvir {\ifmmode r_{\rm vir} \else $r_{\rm vir}$\fi}
\def \rbh {\ifmmode r_{\rm b,pri} \else $r_{\rm b,pri}$ \fi} 
\def \rbs {\ifmmode r_{\rm b,sat} \else $r_{\rm b,sat}$ \fi}

\title[Dark matter and first galaxies]{Clues to the nature of dark matter from first galaxies} 
\author[B. K. Stoychev et al.]{Boyan K. Stoychev$^{1}$,
Keri L. Dixon$^{1}$\thanks{k.dixon@nyu.edu}, 
Andrea V. Macci\`o$^{1,2}$, 
Marvin Blank$^{1,3}$, \and
Aaron A. Dutton$^1$ \\ \\
$^{1}$New York University Abu Dhabi, PO Box 129188, Saadiyat Island, Abu Dhabi, United Arab Emirates\\
 $^{2}$Max Planck Institute f\"{u}r Astronomie, K\"{o}nigstuhl 17, D-69117 Heidelberg, Germany\\
 $^{3}$Institut f\"{u}r Theoretische Physik und Astrophysik, Christian-Albrechts-Universit\"at zu Kiel, Leibnizstr. 15, D-24118 Kiel, Germany}

\begin{document}

\maketitle

\label{firstpage}

\begin{abstract}

We use thirty-eight high-resolution simulations of galaxy formation between redshift 10 and 5 to study 
the impact of a 3~keV warm dark matter (WDM) candidate on the high-redshift Universe.
We focus our attention on the stellar mass function and the global star formation rate and consider the consequences for reionization, namely
the neutral hydrogen fraction evolution and the electron scattering optical depth.
We find that three different effects contribute to differentiate warm and cold dark matter (CDM)  predictions:
WDM suppresses the number of haloes with mass less than few $10^9$~M$_{\odot}$; at a fixed halo mass, WDM produces fewer stars than CDM; and finally at halo masses below
$10^9$~M$_{\odot}$, WDM has a larger fraction of dark haloes than CDM post-reionization.
These three effects combine to produce a lower stellar mass function in WDM for galaxies with stellar
masses at and below $~10^7$~M$_{\odot}$. For $z > 7$, the global star formation density is lower
by a factor of two in the WDM scenario, and for a fixed escape fraction, the fraction of neutral hydrogen is
higher by 0.3 at $z \sim 6$. This latter quantity can be partially reconciled with CDM and observations only by increasing the
escape fraction from 23~per~cent to 34~per~cent.  Overall, our study shows that galaxy formation simulations at high redshift 
are a key tool to differentiate between dark matter candidates given a model for baryonic physics.

\end{abstract}


\section{Introduction}

In the next several years, we expect a large amount of data on the high-redshift ($z>6$) Universe to become available
to the scientific community. Facilities like Atacama Large Millimetre/sub-millimetre Array (ALMA) and 
\textit{James Webb Space Telescope} (\textit{JWST})
will open a completly new observational window 
on the first billion years of the life of our Cosmos. These data will help us understand the early phases of galaxy formation, but they might also guide us towards a better understanding of the dark side of our Universe,
as for example the nature of dark matter.

The current leading model for dark matter is based on a Cold candidate (CDM), 
initially motivated by the possible existence of Weakly Interactive Massive Particles
(WIMPs) predicted by some extensions of the Standard Particle Models \citep{Bertone2005}.
Such candidates should be in the mass range of several GeV \citep{Bergstrom2012} and 
have been extensively searched for in underground labs, but with very little
success to date \citep[see][for a recent review]{Roszkowski2017}. The lack of direct detection of a possible CDM particles raises the question if the dark matter
particle might have a much lower mass than initially thought, and possibly lie in the keV range, 
being what is usually dubbed Warm Dark Matter (WDM).

WDM was initially put forward as a possible solution of the so called `small-scale crisis'
of CDM \cite[e.g][]{Moore1994,Klypin1999, Moore1999,Colin2000}. More recently, there has been a mounting
evidence that the solution of these problems most likely lies in the baryonic sector.
Recent simulations and semi-analytical models have been able to reconcile a CDM universe
with observations both on the number of galactic satellites \citep{Bullock2000, Maccio2010, Sawala2016,
Buck2018} and on the dark matter distribution at small scales 
by simply providing a more accurate treatment of galaxy formation and baryonic physics
within a CDM model \citep{Governato2010, Zolotov2012,  DiCintio2014a, Onorbe2015, Frings2017}.

Moreover, current limits from the Lyman-$\alpha$ (Ly$~\alpha$) forest on the matter power spectrum \citep{Irsic2017,Yeche2017} have
set very stringent limits on the mass of a possible (thermal) WDM candidate, which is now constrained to
$m_{\nu} >3.0$~keV. With such a mass, a WDM candidate will be practically
indistinguishable from a cold one regarding the dark matter distribution on small scales \citep{Colin2008,Maccio2012b, Shao2013, Herpich2014}
or the number of satellites \citep{Maccio2010b, Lovell2016}. The similarity of currently allowed WDM models to CDM on small scales calls for a new venue
on where to look for more clues on the actual mass of the dark matter candidate, and this venue
can lie in the high-redshift Universe. 

In WDM, due to the lack of power on small scales, structure formation is delayed with respect
to a cold scenario. At present time, the effect of such an initial delay has been erased by the nonlinearity 
of galaxy formation, but it can still present itself in the very early stages of the Universe, when the first galaxies were formed. This signature can manifest itself in the formation of first stars \citep[e.g.][]{Gao2007}, or first galaxies 
\citep[e.g.][]{Dayal2015}.
Recently, \cite{Corasaniti2017} have used measurements of the galaxy luminosity function at $z =$ 6, 7, and 8 to derive constraints on WDM. 
They have combined very high resolution $N$-body (pure gravity) simulations with an empirical approach based on halo abundance matching
to derive predictions for the luminosity function in different dark matter models.
They have obtained a lower limit on the WDM thermal relic particle mass of 1.5~keV at a 2$\sigma$ level. 
This study relies on dark matter only simulations that might miss some
of the nonlinear effects between the nature of dark matter and galaxy formation, which is a possible limitation.

The first galaxies are also thought to be the main source of the ionizing photons responsible for 
cosmic reionization, which may then provide another constraint to the nature of dark matter. Many recent studies have studied the connection between WDM and reionization \citep[e.g.][]{Lapi2015, Tan2016, Lopez2017, Das2018}. In particular, \citet{Carucci2019} extend the work of \cite{Corasaniti2017} to derive the requisite escape fractions of ionizing photons to match recent observations, but given the uncertain nature of this quantity, the results largely independent of underlying dark matter scenario. Two studies employing semi-analytic methods to study WDM in the context of high redshift and reionization also find that with a dark matter scenario consistent with observational constraints the ionizing population is shifted to lower halo masses, but that distinguishing CDM and WDM with current observations would be difficult \citep{Bose2016, Dayal2017}.
Furthermore, \cite{Villanueva2018} have used hydrodynamical simulations in a box of 20~$h^{-1}$~Mpc to
derive constraints on WDM from galaxy luminosity functions, 
the ionization history, and the Gunn-Peterson effect. 
They have concluded that while some effects are indeed present, until the modelling of baryonic effects (star formation
and feedback mainly) are constrained better, any conclusions on the nature of dark matter derived from
reionization observables remain model-dependent. One possible way to anchor the parameters describing the baryonic physics (besides having better constraints from observations)
is to use observations at one redshift (usually $z=0$) as a constraint and then check predictions at a much higher redshift.

In this study, we attempt to connect the high- and low-redshift Universe via hydrodynamical simulations of galaxy formation.
We want to extend to the high-redshift Universe ($z>6$) 
the baryonic physics used in the NIHAO project \citep{Wang2015}, which has proven very successful in reproducing
key observations in the low-redshift Universe ($0<z<2$), as for example the stellar mass - halo mass relation \citep{Wang2015}, the 
local velocity function \citep{Maccio2016}, and galaxies scaling relations \citep{Dutton2017}.

We use 19 high-resolution, zoom-in simulations in each cosmology (for a total of 38) to accurately sample the stellar mass function in the redshift range $5<z<10$,
and we study the impact of a 3~keV dark matter candidate on several current and future observables, like the stellar mass function, 
the global star formation rate, the hydrogen neutral fraction, and the Cosmic Microwave Background (CMB) optical depth.

The paper is organized as follows. In Section~\ref{sec:setup}, we present the set up of our galaxy formation simulations, 
in Section~\ref{sec:results}, we present our results and forecasts for current and future observations. Finally in 
Section~\ref{sec:concl}, we discuss the implications or our results and present our conclusions.

\section{Simulations}
\label{sec:setup}

All simulations herein were ran in a flat $\rm \Lambda CDM$/$\rm \Lambda WDM$ cosmology, with parameters from the Planck Collaboration et al. \citep{Planck2014}: Hubble Parameter $H_{0} = 67.1$ km s$^{-1}$ Mpc$^{-1}$, matter density $\Omega_{\rm m} = 0.3175$, radiation density $\Omega_{\rm r} = 0.00008$, dark energy density $\Omega_{\Lambda} = 1 - \Omega_{\rm m} - \Omega_{\rm r} = 0.6824$, power spectrum normalization $\sigma_{\rm s} = 0.8344$, and power spectrum slope $n = 0.9624$.

\subsection{Initial Conditions}
\label{ssec:IC}

Initial conditions for both the $N$-body and hydrodynamical, zoom-in simulations of galaxy formation were generated using a modified version of the {\sc \small grafic2} package \citep{Bertschinger2001, Penzo2014}. For WDM, the process is identical to CDM except for a modification to the power spectrum. The same random seed was used to generate all pairs of CDM/WDM 3~keV initial conditions. 

WDM power spectra were computed with the use of a relative transfer function, as in \cite{Bode2001}:

\begin{equation}
	P_{\rm WDM}(k) = [T_{\rm WDM}(k)]^2 P_{\rm CDM}(k),
\end{equation}
where
\begin{equation}
	T_{\rm WDM}(k) = \left[1+(\alpha k)^{2\nu}\right]^{-5/\nu}.
\end{equation}
Here, $\nu = 1.12$ \citep{Viel2005} and $\alpha$ is the length scale of the break in the WDM power spectrum:

\begin{equation}
	\alpha = 0.049\left(\frac{m_{\rm x}}{1 {\rm keV}}\right)^{-1.11}\left(\frac{\Omega_{\rm x}}{0.25}\right)^{0.11}
	\left(\frac{h}{0.7}\right)^{1.22} h^{-1} {\rm Mpc},
\end{equation}
where $m_{\rm x}$ is the mass of the WDM particle and $\Omega_{\rm x}$ the density.

\subsection{$N$-body Simulations}

\begin{table}
\centering
\begin{tabular}{ccccc}
\hline
\hline
Box & Size  & N & m$_{\rm dm}$  & $\epsilon_{\rm dm}$ \\
\hline
  & Mpc  &   & M$_{\rm \odot}$ & kpc \\
\hline
1  &   $5$     & $250^3$  &  $1.00 \times 10^6$ &  $0.50$ \\
2  &   $15$    & $250^3$  &  $2.70 \times 10^7$ &  $1.50$ \\ 
3  &   $20$    & $500^3$  &  $8.00 \times 10^6$ &  $1.00$ \\
4  &   $40$    & $600^3$  &  $3.70 \times 10^7$ &  $1.67$ \\   
5  &   $100$   & $500^3$  &  $1.00 \times 10^9$ &  $5.00$ \\   
6  &   $200$   & $500^3$  &  $8.00 \times 10^9$ &  $10.0$ \\  
\hline
\hline
\end{tabular}
\caption{Size and resolution of the 6 $N$-body simulation boxes.}
\label{tab:box}
\end{table}

First, we run a series of $N$-body simulations that only include dark matter, detailed in Table~\ref{tab:box}. We used six boxes that range from 5 to 200~Mpc length on a side with varying particle number. Softening was set to 1/40 of the intra-particle distance on the initial conditions grid. From this set of simulations, we are then able to sample a wide range of halo masses at $z= 5 - 10$.

\subsection{Hydrodynamical Simulations}
From the dark matter-only simulations, 19 haloes were chosen to be re-simulated at higher resolution with baryons included. Table~\ref{tab:res} shows the five resolution levels used in our simulations. These levels were chosen to have approximately one million particles
inside the virial radius of the galaxy across the whole range of halo masses.

\begin{table}
\centering
\begin{tabular}{ccccc}
\hline
\hline
Level   & $m_{\rm DM}$ & $m_{\rm gas}$  & $\epsilon_{\rm DM}$ & $\epsilon_{\rm gas}$ \\
\hline
   & M$_{\rm \odot}$ & M$_{\rm \odot}$ & pc & pc \\
\hline
1  & $1.95 \times 10^3$  &  $3.56 \times 10^2$ &  $63$  &  $27$ \\
2  & $1.56 \times 10^4$  &  $2.85 \times 10^3$ &  $125$  &  $53$ \\
3  & $1.25 \times 10^5$  &  $2.28 \times 10^4$ &  $250$  &  $107$ \\
4  & $4.21 \times 10^5$  &  $7.70 \times 10^4$ &  $375$  &  $160$ \\
5  & $5.78 \times 10^5$  &  $1.06 \times 10^5$ &  $417$  &  $178$ \\
\hline
\hline
\end{tabular}
\caption{Resolution levels used in our zoom-in simulations.}
\label{tab:res}
\end{table}

All simulations were performed with the SPH code {\sc \small gasoline} \citep{Wadsley2004, Wadsley2017}. The code was setup in the framework of the NIHAO project \citep{Wang2015}, including metal cooling, chemical enrichment, star formation, and feedback from massive stars and supernovae (SN).

Stars are formed from gas cooler than $T$ = 15,000~K and denser than
$n_{\rm th}=10{\rm ~cm}^{-3}$. The star formation efficiency used was $c_\star$ = 0.1. 
Cooling via hydrogen, helium, and various metal-lines is included as in \citet{Shen2010}, including photoionization and heating from ultraviolet (UV) background in \citet{Haardt2012}.

SN feedback is implemented using the blastwave approach as
described  in \citet{Stinson2013}, which relies on delaying the cooling
of nearby particles to a SN event. We also include what we dubbed Early
Stellar Feedback: we inject 13~per~cent of the UV
luminosity of the stars as thermal energy before any SN events take place without
disabling the cooling \citep[see ][for more details]{Stinson2013}. 

NIHAO galaxies have been shown to be consistent with a wide range of low-redshift galaxy properties, such as the stellar mass - halo mass relation and star formation rates \citep{Wang2015}; stellar disk kinematics \citep{Obreja2016}; cold and hot gas content \citep{Stinson2015, Wang2016, Gutcke2016}; and resolve the too-big-to-fail problem \citep{Dutton2016}. We are therefore confident that the simulations presented herein can be used as plausible tools to study the effect of a WDM cosmology on the high-redshift Universe.

Haloes in all simulations were identified using the MPI+OpenMP
hybrid Amiga Halo Finder \footnote{http://popia.ft.uam.es/AMIGA}
\citep[\textsc{\small AHF};][]{ahf}. The virial masses of the
haloes are defined as the mass within a sphere containing
$\Delta = 200$ times the cosmic critical matter density. The virial
(total) mass is denoted as $M_{200}$, and $M_{\star}$ indicates the total stellar mass within the virial radius. 

Beyond the 19 target galaxies, we include any galaxy within our zoom simulations that has greater than 10,000 particles. We also only consider central haloes, i.e. not satellites, and require no pollution from larger low-resolution dark matter particles, i.e. any particle larger than the lowest level in Table~\ref{tab:res}. The result is of order 100 galaxies at $z = 5$, and tens of galaxies at $z = 10$.
Simulation analysis was done using the Python package {\sc \small pynbody} \citep{pynbody}. All parameter fits were performed using Markov Chain Monte Carlo (MCMC) via the Python package {\sc \small emcee} \citep{emcee}.

\section{Results}
\label{sec:results}

In the following, we will present the results of our $N$-body and galaxy simulations in CDM and WDM cosmologies at $z = 5-10$. In all plots, CDM results are shown in blue with circles, while 3~keV WDM results are shown in red with squares. Section \ref{sec:hmf} shows the halo mass functions obtained from the $N$-body simulations. In Section \ref{sec:smhm}, we find the stellar mass - halo mass relation obtained from our simulations. Section~\ref{sec:frac} shows our results for the fraction of galaxies that form stars as a function of halo mass. Section~\ref{sec:smf} shows the stellar mass functions obtained by convolving all prior results. Finally in Section~\ref{sec:reion}, we compare our results for the reionization history and CMB optical depth to observational constraints.

\subsection{Halo Mass Function}
\label{sec:hmf}

The $N$-body boxes were used to construct and fit halo mass functions. We used a five-parameter, modified Schechter function \citep{Schecter1976}:
\begin{equation}
\label{eq:hmf}
\log\left(\frac{dN}{d\log M}\right) = A-B \log M - C{\rm e}^{\left(M_{0}-\log M\right)^{\alpha}},
\end{equation}
where $A$, $B$, $C$, $\alpha$, and $M_0$ are fitting parameters that are given in Appendix~\ref{app:hmf} and $M$ represents $M_{200}/\Msun$. The original Schechter function struggled to fit the WDM results, so we adopted this modified version for both cosmologies. The uncertainty for each halo mass bin is considered Poissonian.

\begin{figure}
\centering
	\includegraphics[width=80mm]{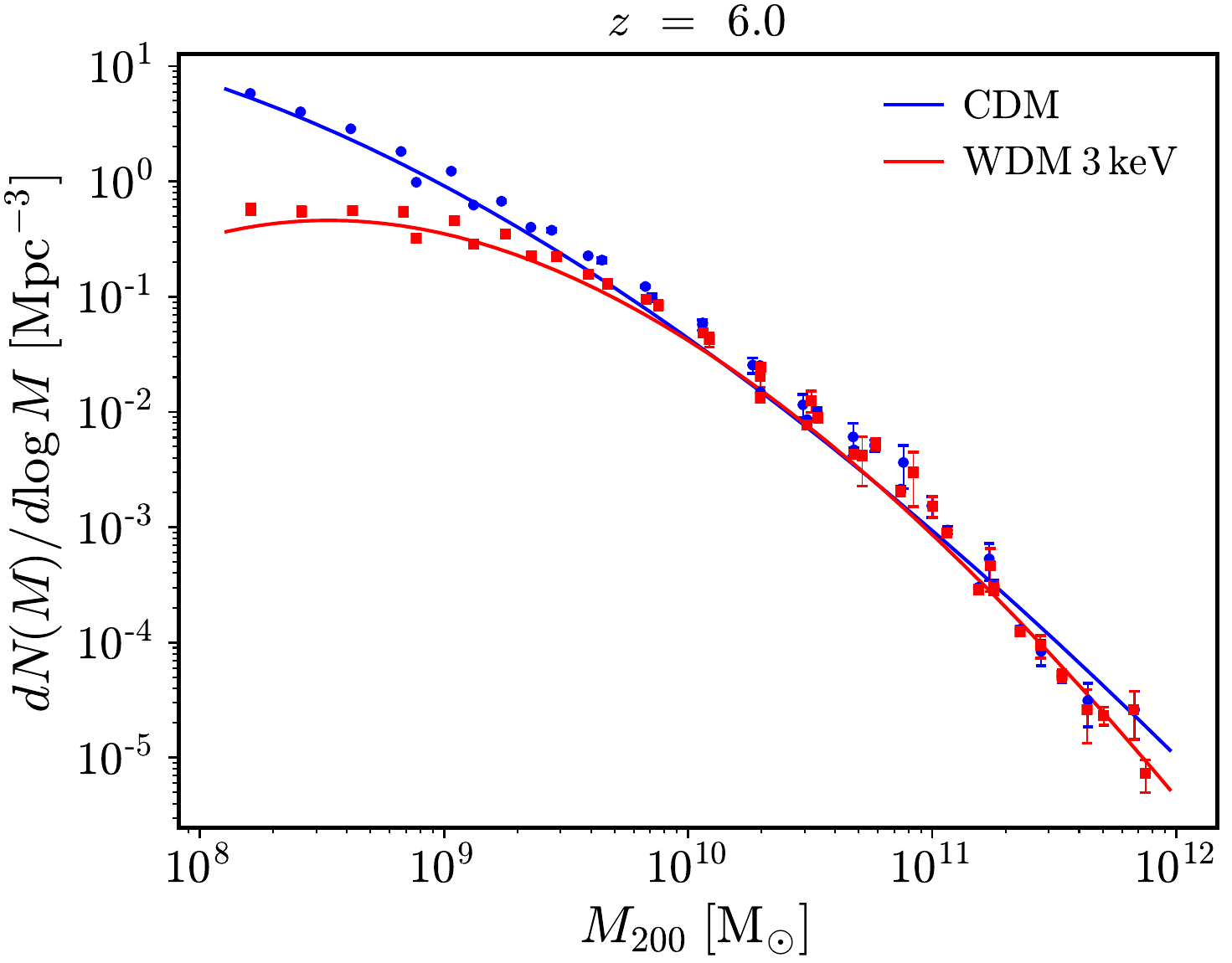}
	\includegraphics[width=80mm]{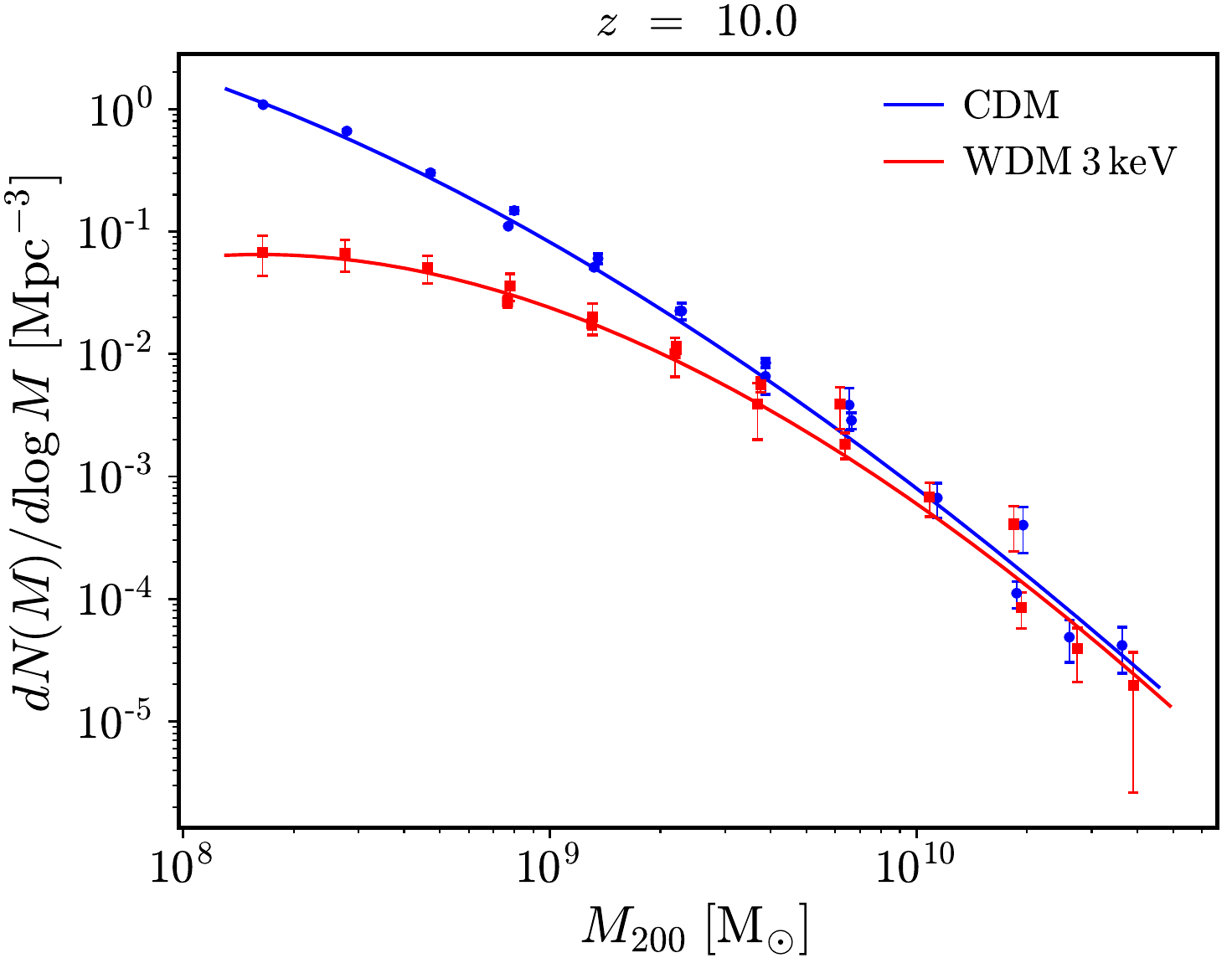}
	\caption{Halo mass function at $z = 6$ (upper panel) and 10 (lower panel). The blue circles and red squares correspond to CDM and WDM results as obtained from the $N$-body simulations, respectively. The error bars represent Poisson errors, reflecting the smaller number of massive haloes at high redshift. The familiar turnover of the halo mass function for WDM at low halo mass is evident in both panels.}
	\label{fig:hmf}
\end{figure}

The halo mass function represents the abundance of haloes at a given mass. As shown in Fig.~\ref{fig:hmf}, the halo mass functions exhibit a suppression of the number density of low-mass haloes in a WDM cosmology, a direct consequence of the lack of power on small scales in the WDM power spectrum. The lines represent the MCMC parameter fit, the values of which are given in Appendix~\ref{app:hmf}. The relations for the two cosmologies follow the same trend at $z = 5-10$. Importantly, the difference between CDM and WDM is larger at higher redshift. At $z = 0$, the halo mass functions would be identical for the two scenarios.

\subsection{Stellar Mass - Halo Mass Relation}
\label{sec:smhm}

\begin{figure}
\centering
	\includegraphics[width=80mm]{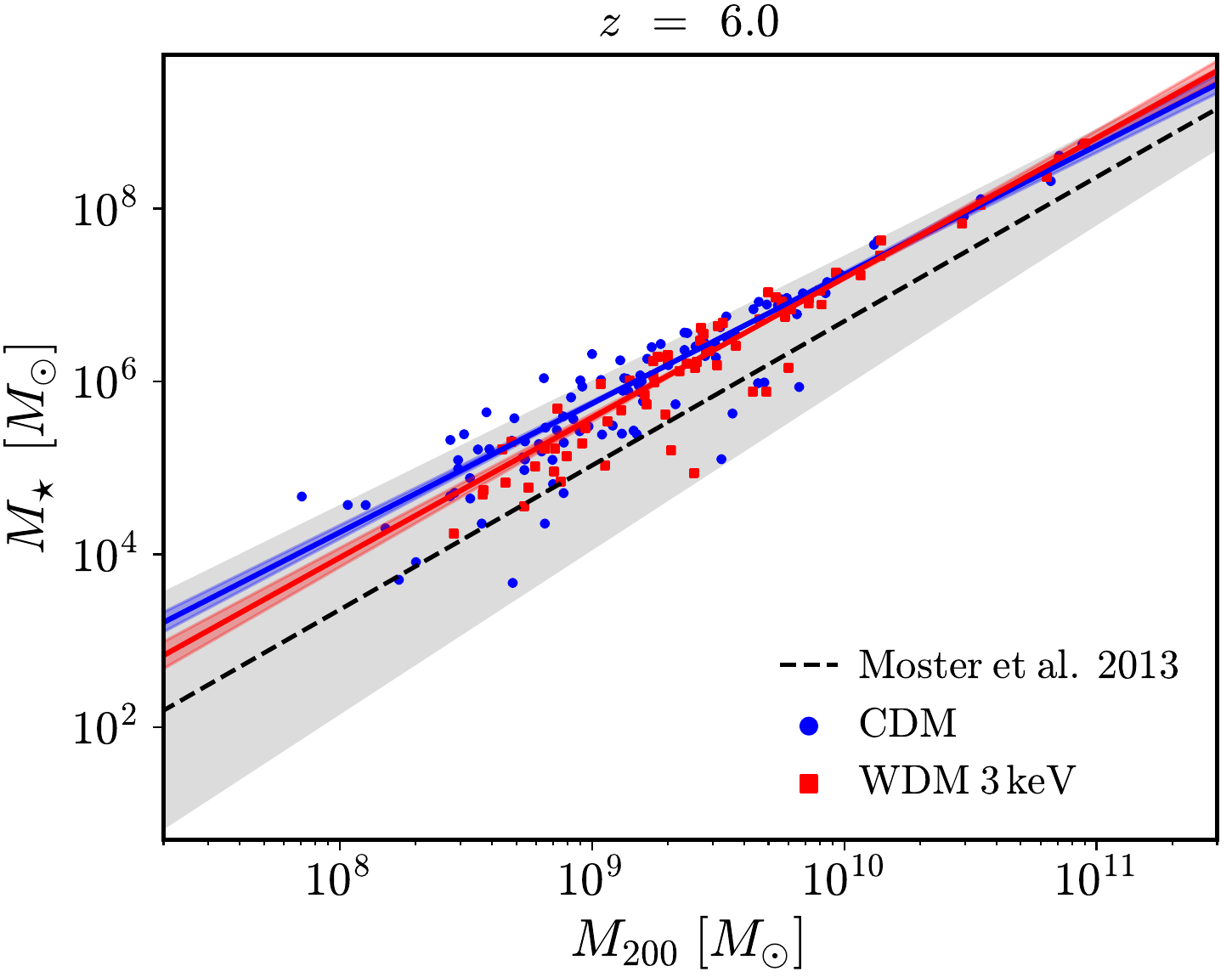}
	\includegraphics[width=80mm]{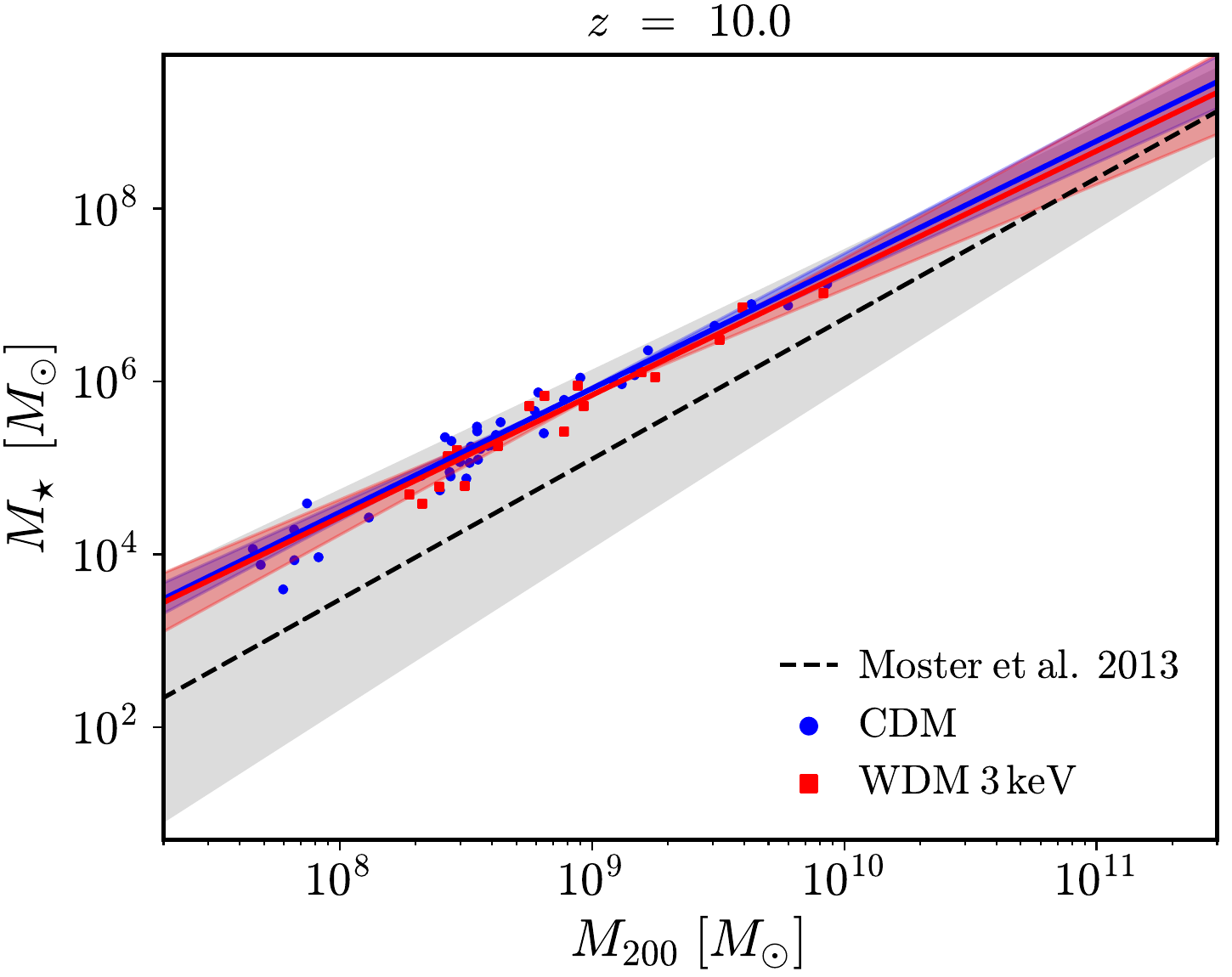}
	\caption{Stellar mass - halo mass relation at $z = 6$ (upper panel) and 10 (lower panel). Each blue circle and red square corresponds to a hydrodynamical zoom-in simulation in CDM and WDM, respectively. The blue and red lines show a simple power-law fit. The dashed line shows the well-known Moster relation with the grey area showing the $1\sigma$ scatter  \citep{Moster2013}. Importantly, this relation is extrapolated to higher redshift and lower masses than the original fit.}
	\label{fig:smhm}
\end{figure}

Fig.~\ref{fig:smhm} shows the stellar mass - halo mass relation obtained from the set of zoom-in simulations run in CDM and WDM cosmologies.\footnote{To be included in Fig.~\ref{fig:smhm} and the fit used in Section~\ref{sec:reion}, we require that the haloes have at least 10 stellar particles.}  Based on this relation, CDM and WDM galaxies cannot be distinguished from each other. While our simulations closely follow the abundance matching relation(s) at lower redshift \citep[see][]{Wang2015}, our predicted relation at these high redshifts consistently lies above the extrapolation from \cite{Moster2013}.\footnote{The original NIHAO simulations of \cite{Wang2015} use the stellar mass within 20 percent of the virial radius, which would bring our results even closer to the relation. At high redshift, this cut is too conservative.} This offset (minor in our simulations) suggests that high-redshift extrapolations of relations calibrated at $z < 4$ should be used with caution. Many studies at high redshift exhibit similar offsets, see e.g. \cite{Rosdahl2018}, but also see \cite{Ma2018}. This figure also indicates that a larger fraction of WDM galaxies remain dark at high redshift compared to CDM, as seen by the smaller number of red points relative to blue. This effect is further explored in the next section.

The solid lines represent a linear fit to the logarithmic data, which is done independently for each cosmology and redshift. The median (1$\sigma$) parameters derived from the MCMC sample generate the solid lines (shaded regions), the values of which are located in Appendix~\ref{app:hmf} These (small) uncertainties are propagated to the stellar mass function results. As shown in Fig.~\ref{fig:smhm}, CDM and WDM behave similarly. We find weak evidence for redshift evolution of the normalization in that lower redshift has a higher value for the intercept of the linear fit, which is consistent with \citet{Ceverino2017}. However, \cite{Ma2018} find no redshift evolution.

\subsection{Dark Fraction}
\label{sec:frac}

Star formation becomes more inefficient at low masses, to the point where no gas is able to collapse to the center and halo and form stars due to the UV background \citep{Gnedin2000}.
Fig.~\ref{fig:smhm} seems to suggest a larger fraction of `dark' galaxies in WDM w.r.t. CDM, given that there are fewer red squares than blue circles. To quantify this effect, we want to compute the fraction of objects able to form stars as a function
of their virial mass. A halo is considered dark if it contains no stellar particles.

\begin{figure}
\centering
	\includegraphics[width=85mm]{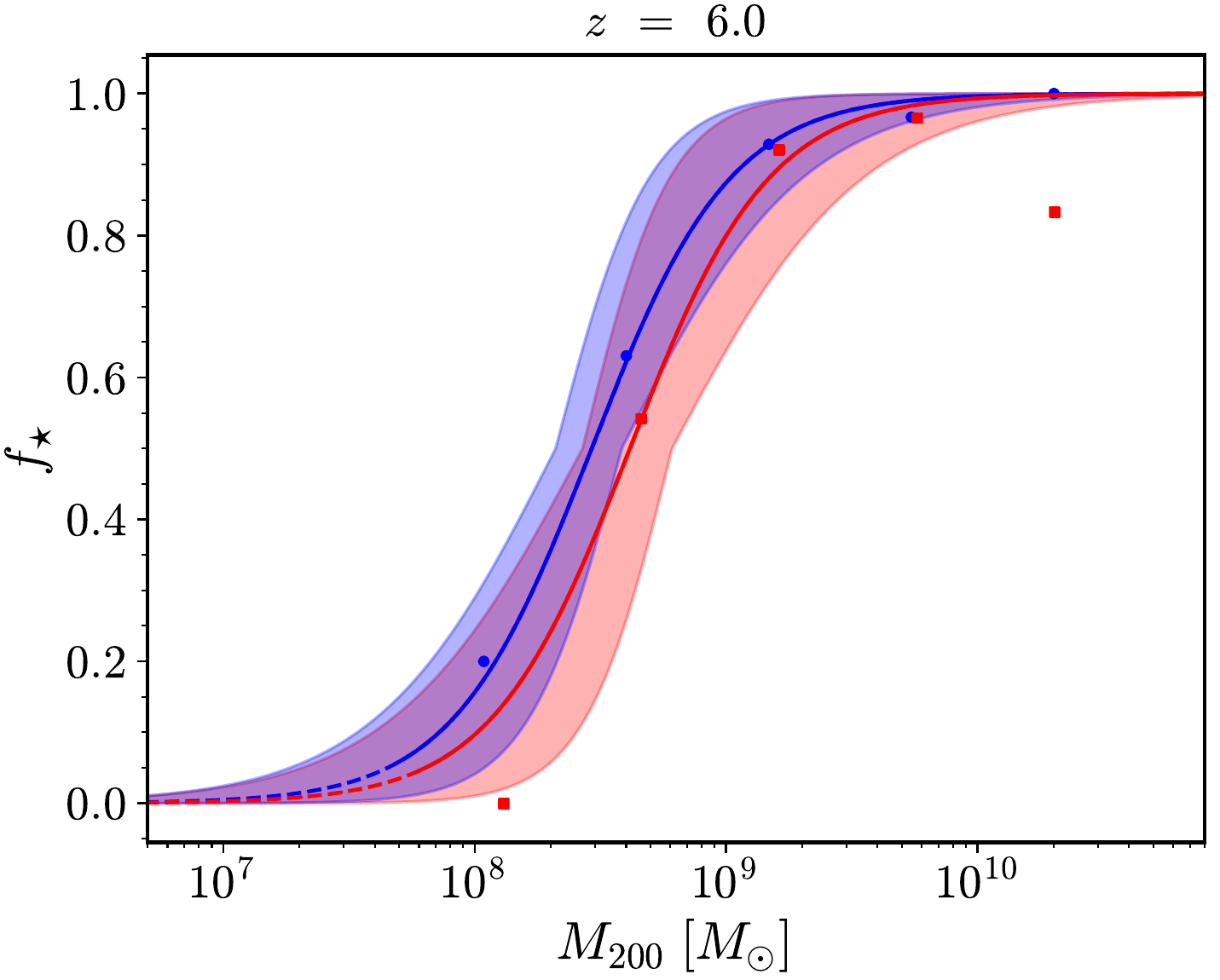}
	\includegraphics[width=85mm]{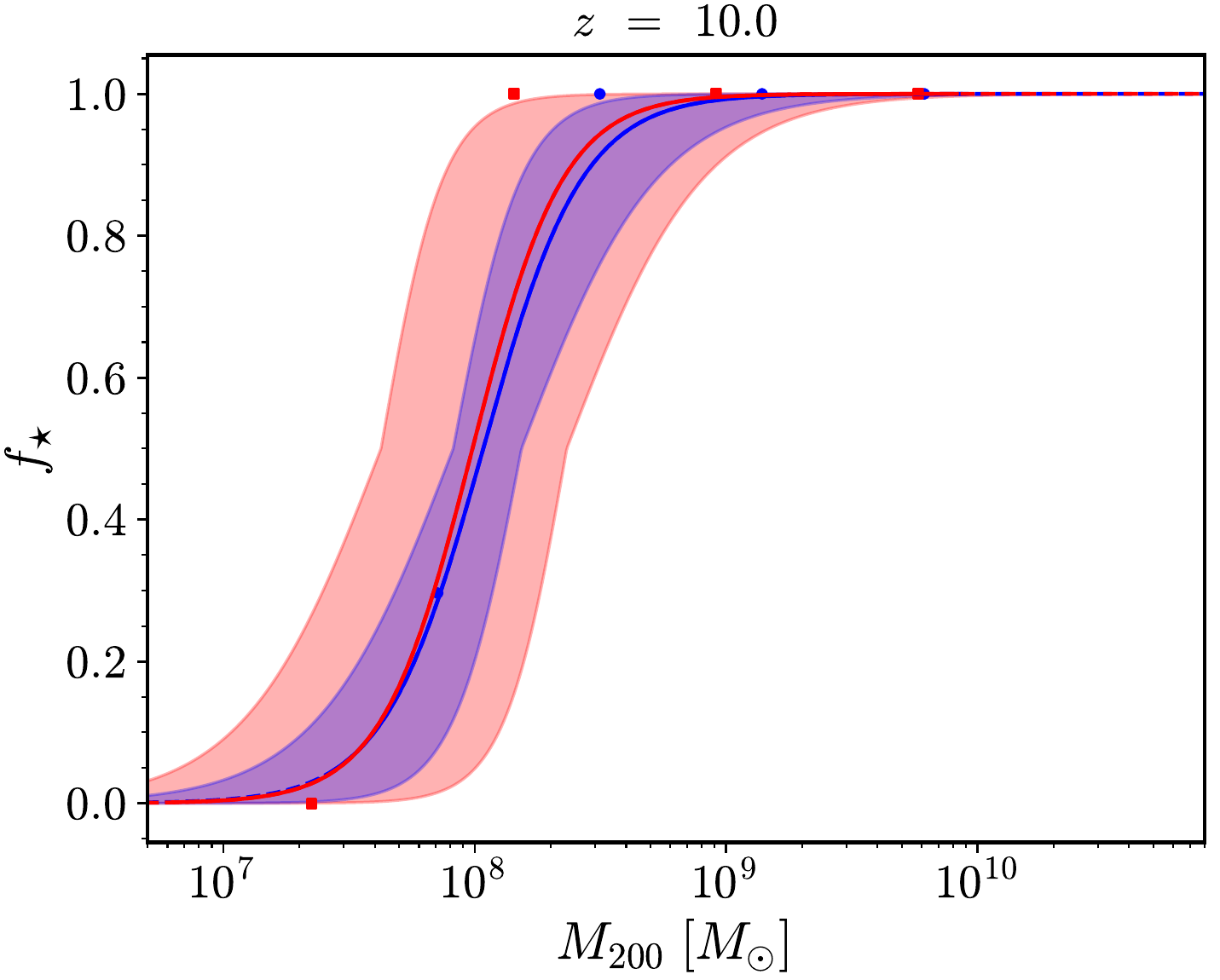}
	\caption{Fraction of galaxies containing stars as a function of their halo mass. The points represent the binned values for CDM (squares) and WDM (circles). The solid lines are the best-fitting from the MCMC sample, and the shaded regions represent the 1$\sigma$ uncertainties on the parameters derived for the sample.}
	\label{fig:frac}
\end{figure}

Results are presented in Fig.~\ref{fig:frac} for $z = 6$ and $z = 10$. The fitting curve is a hyperbolic tangent two-parameter function 

\begin{equation}
f_{\star} = \frac{1+\tanh\left[\, \beta ({\rm log}M-M_{1})\right]}{2},
\label{eq:fstar}
\end{equation}
where $\beta$ and $M_1$ are parameters and $M$ once again represents $M_{200}/\Msun$. The points represent the histogram of the $f_{\star} = 1 - N_{\rm dark}/N_{\rm bin}$, where $N_{\rm dark}$ are the number of dark haloes and $N_{\rm bin}$ are the total number of haloes in the bin. When fitting this curve, the uncertainties on $f_{\star}$ are derived from the number of haloes in each bin. The resultant median (1$\sigma$) uncertainties of the fitted parameters are generated from the MCMC sample and are represented by the solid line (shaded region), all values are found in Appendix~\ref{app:hmf}. We find weak evidence that WDM exhibits a steeper dark fraction fraction transition, which is consistent with \cite{Maccio2019}. 

In these simulations, reionization occurs $z \gtrsim 9$, and the shift of minimum halo mass for star formation is evident. At earlier times (higher redshift), the $M_{200}$ at which stars form is lower due to a lower UV background. At $z > 8$, WDM has a lower minimum mass for star formation than CDM, and vice versa for lower redshift. Another interesting result is that only approaching $M_{200} > 5\times10^{10} \Msun$ guarantees that stars will form in galaxies at this redshift. The main caveat for these observations is that the uncertainties for $f_{\star}$ are large given our small halo numbers, and distinguishing with confidence between the two cosmologies is beyond our statistical reach.

\subsection{Stellar Mass Function}
\label{sec:smf}

\begin{figure*}
\centering
	\includegraphics[width=85mm]{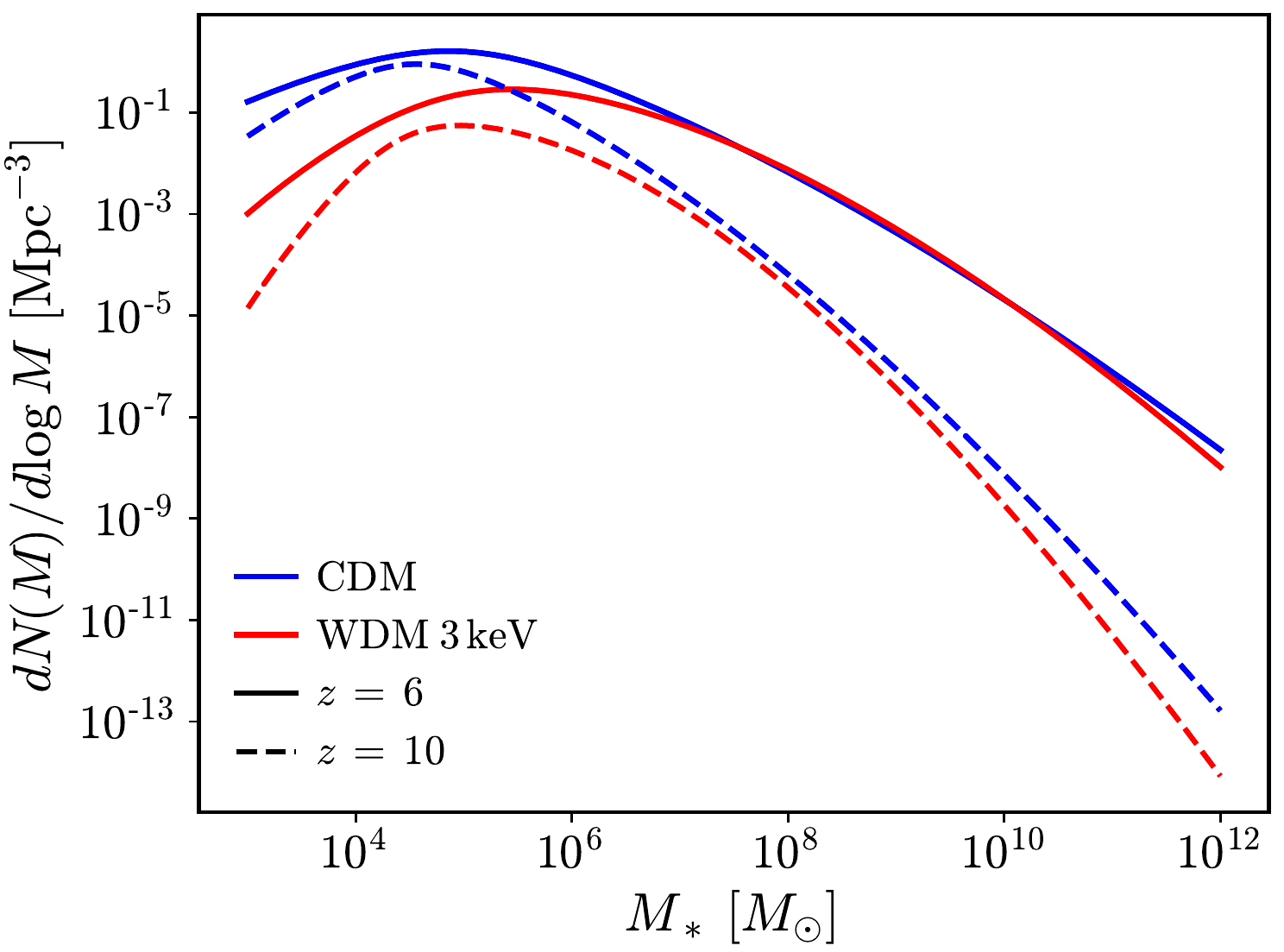}
	\includegraphics[width=85mm]{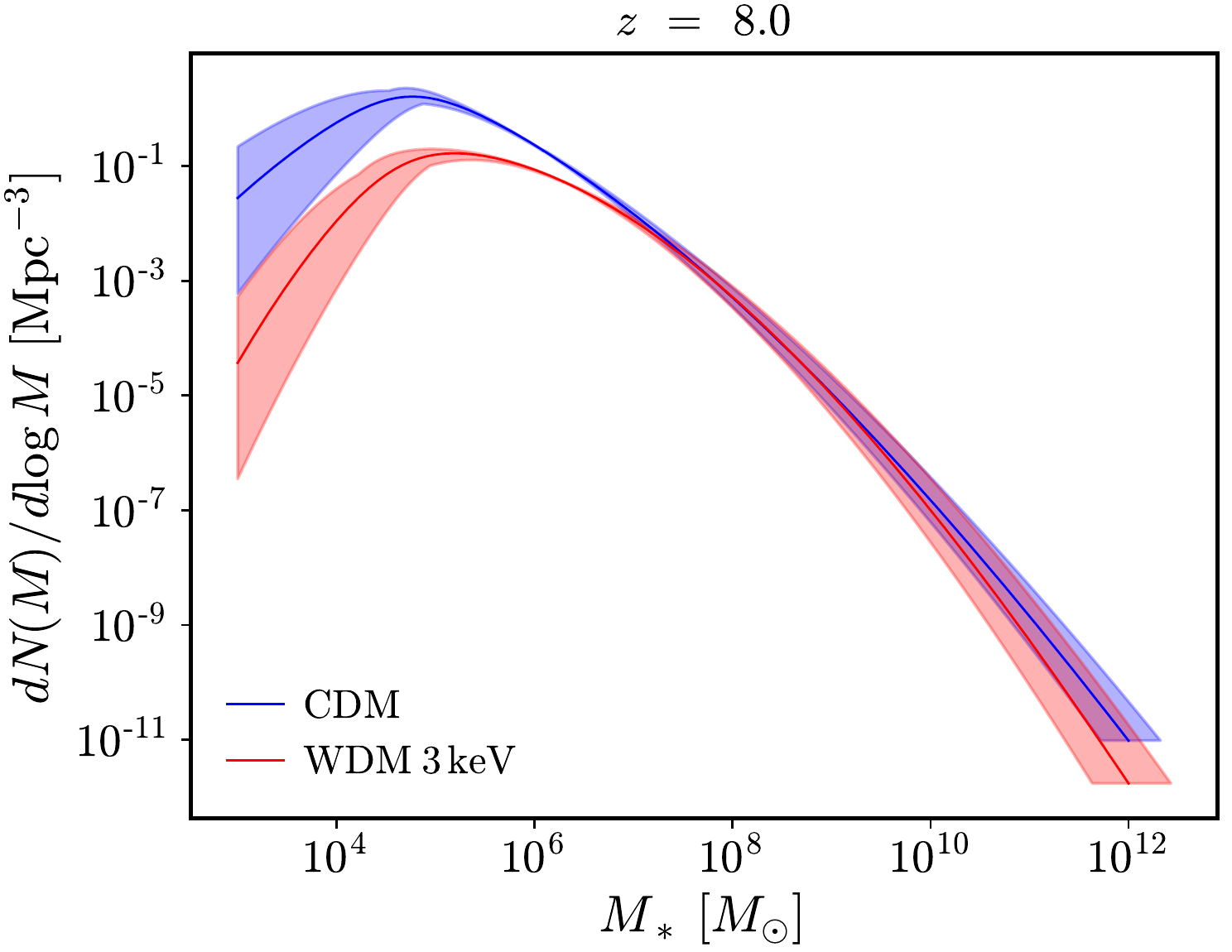}
	\caption{Left: Stellar mass functions at $z$ = 6 (solid) and 10 (dashed) in CDM (blue) and WDM 3~keV (red) cosmologies. The WDM turnover at low $M_{\star}$ is more pronounced than the halo mass function case. Right: The stellar mass functions and 1$\sigma$ errors for both cosmologies at $z$ = 8. The shaded region represents the 1$\sigma$ uncertainty as estimated by propagating the uncertainty from the MCMC fitting of the $M_{\star}$-$M_{200}$ relation (dominates at high masses) and $f_{\star}$ (dominates at low masses).  }
	\label{fig:smf}
\end{figure*}

The next step is to convolve the results from previous sections in order to obtain the stellar mass function. We started from the fitted stellar mass - halo mass relation, including the uncertainty in the parameters, 
for our set of galaxy simulations (Fig.~\ref{fig:smhm}) to convert the halo masses in Fig.~\ref{fig:hmf} into stellar masses. We then multiplied by the fraction, including the uncertainty of the fit,  
of galaxies that actually formed stars (Fig.~\ref{fig:frac}) to obtain the differential number density of galaxies as a function of their stellar mass, which is shown in 
Fig.~\ref{fig:smf}. The solid line represents the best-fitting (and median) parameters and the shaded region is the propagated uncertainties, where the $M_{\star}$-$M_{200}$ relation dominates at high stellar masses (few high-mass haloes) and $f_{\star}$ dominates at low stellar masses (suppression of star formation becomes important). The $M_{\star}$-$M_{200}$ uncertainty is reflected the stellar massCeverino2017 uncertainty ($x$ axis), and $f_{\star}$ drives the uncertainty in the number density ($y$ axis).

As expected, the stellar mass functions substantially differ only at low masses, while the small differences at high masses are due to cosmic variance and low number statistics.
WDM predicts fewer low-mass galaxies than CDM, with a larger difference at high redshift.
For our choice of warm candidate mass (3~keV), the relative difference between CDM and WDM is only a factor of a few stellar masses around $10^6$~\Msun \citep[see also][]{Villanueva2018},
below what is currently observable \citep{Bouwens2015, Bouwens2017, Ceverino2017}.
On the other hand, our results are in agreement with \cite{Corasaniti2017}, who found a lower limit of 2~keV from analysis of the luminosity function. 
Future observations and facilities might improve this limit (e.g. \textit{JWST}), but then a very careful understanding of the baryonic physics implemented in the simulations
will be needed, as discussed in \cite{Villanueva2018}.

\subsection{Star Formation Rates}
\label{sec:sfr}

\begin{figure}
\centering
	\includegraphics[width=80mm]{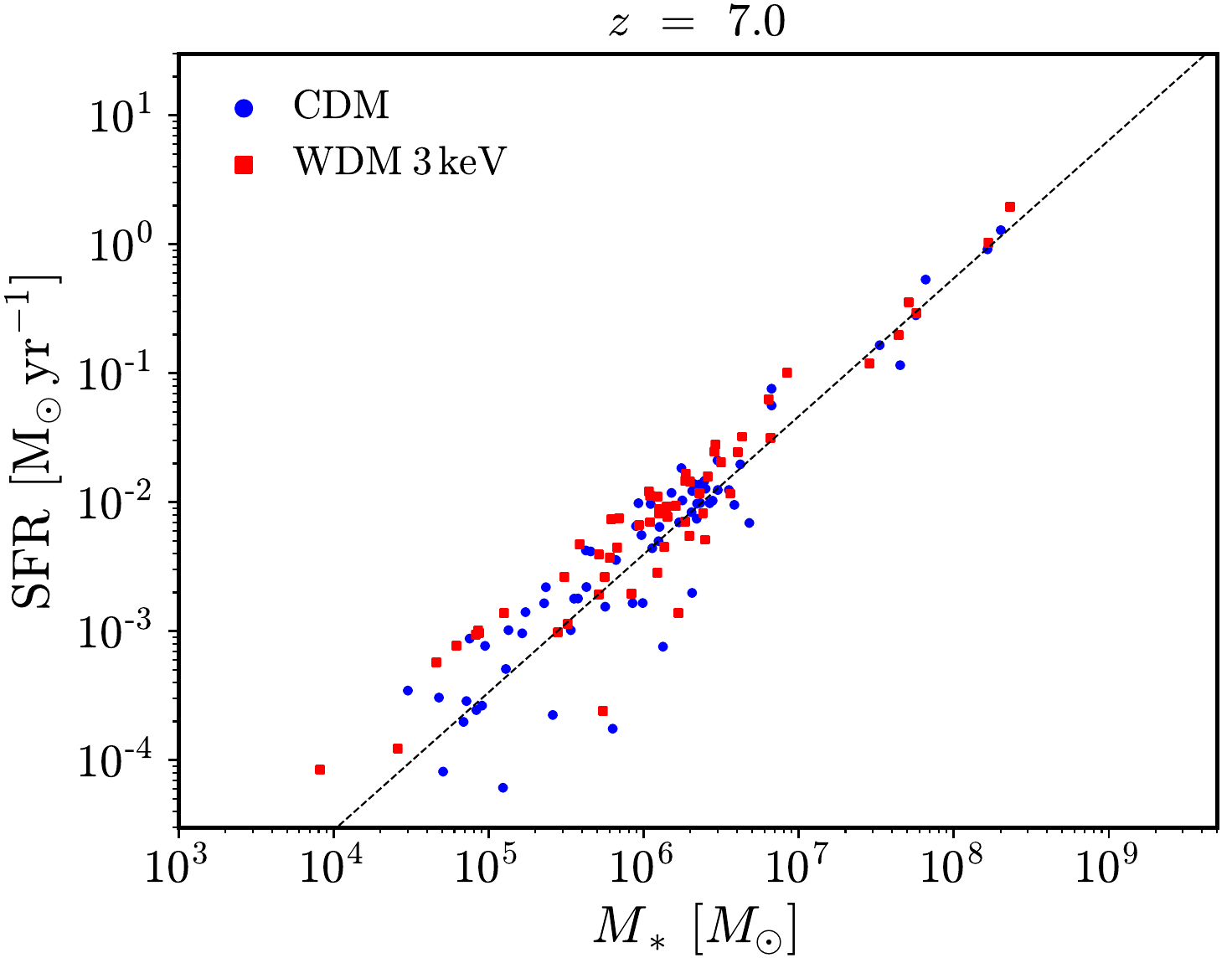}
	\caption{Star formation rate - stellar mass relation in CDM (blue circles) and WDM (red squares) at $z = 7$. Here, the SFR is averaged over the previous 100~Myr. The dashed line represents the linear (in log space) fit to the combined data set of CDM and WDM, as there is no significant difference in the relation between the two cosmologies.}
	\label{fig:smsfr}
\end{figure}

The star formation rate (SFR) of a galaxy is not only of interest on its own, the reionization history of the Universe is crucially dependent on this quantity. First, we compute the SFR-$M_{\star}$ relation for our galaxies as a function of redshift, where the SFR includes star formation occurring within the past 100~Myr at the redshift in question.  As shown in Fig.~\ref{fig:smsfr}, CDM and WDM follow the same relation at $z = 7$, and the same is true for lower and higher redshifts. With this in mind, we fit this relation with a single power law for both models at each redshift (shown by the dashed line). The usual MCMC procedure was performed with each SFR value weighted by the number of stellar particles that formed in the past 100~Myr. The best-fitting slope and intercept values\footnote{As a check, the $1\sigma$ uncertainties were calculated as detailed in previous relation fittings, but the resultant values were very small and would be entirely subdominant to the uncertainties in $M_{\star}$-$M_{200}$ and $f_{\star}$. As such, the uncertainty in this relation is not presented nor propagated to the stellar mass functions.} are found in Appendix~\ref{app:hmf} and are similar to those found in \citet{Ma2018}. Contrary to \citet{Ma2018}, we find evidence of redshift evolution in the intercept or normalization, though no evolution in the slope. Therefore, at lower redshift, a lower SFR is expected for the same stellar mass.

\begin{figure}
 \centering
 	\includegraphics[width=85mm]{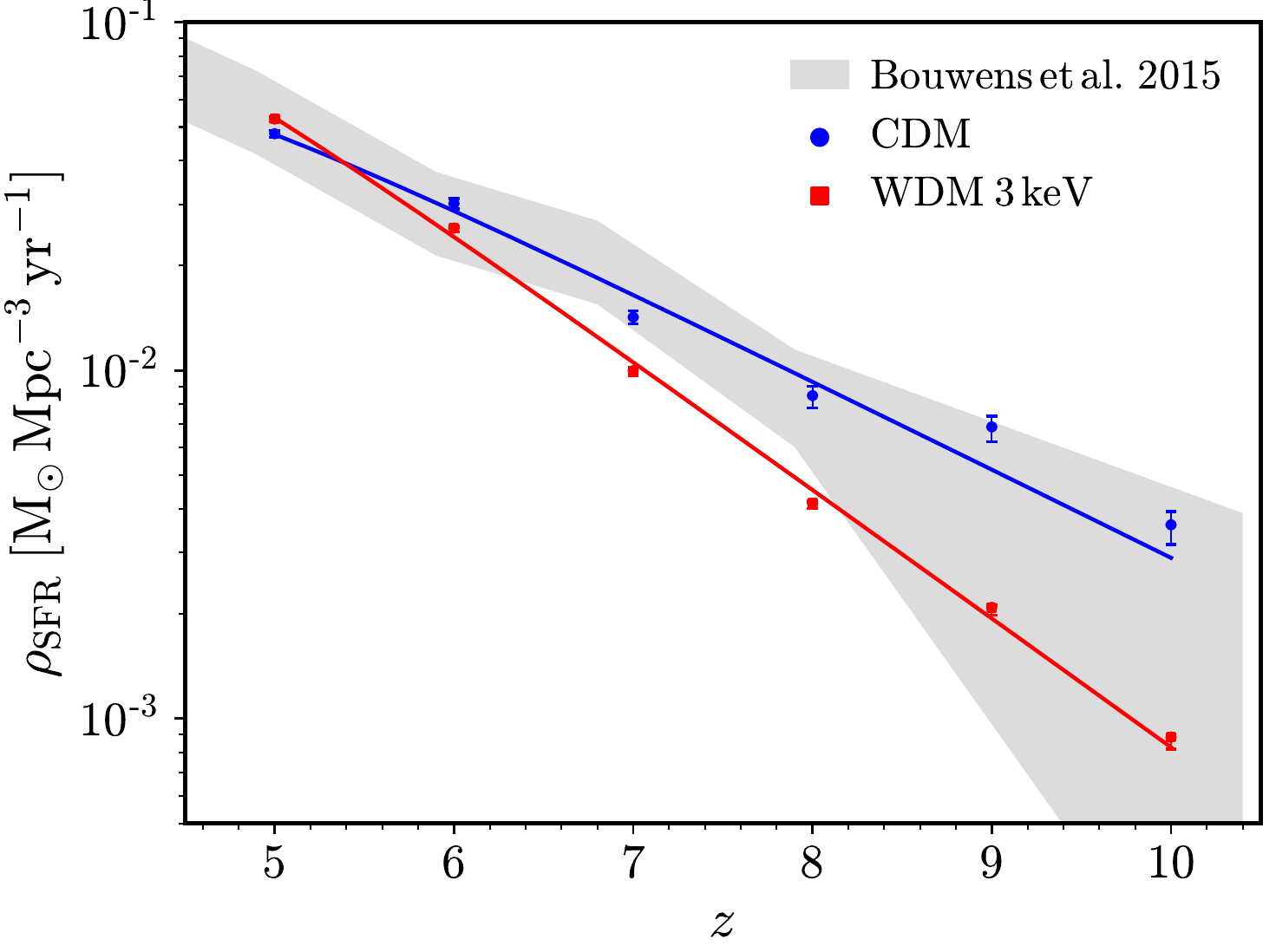}
 	\caption{Global star formation rate density as a function of redshift. The shaded region encompasses the data with and without dust correction. The WDM scenario has lower $\rho_{\rm SFR}$ than CDM at higher redshift, but by $z = 6$, the two cosmologies roughly reach parity. }
 	\label{fig:sfr}
 \end{figure}

We then combine our stellar mass functions with the fits of SFR versus $M_{\star}$  to obtain an integrated global SFR density, $\rho_{\rm SFR}$, as a function of redshift (Fig.~\ref{fig:sfr}). 
Stellar mass functions were integrated between $10^3 - 10^{12}~\Msun$, and the results are insensitive to order-of-magnitude changes in these limits. The values of $\rho_{\rm SFR}$ obtained for $z = 5-10$ were fitted following:

\begin{equation}
\label{eq:sfrdens}
\log\left(\frac{\rho_{\rm SFR}(z)}{\rm M_{\odot} \, Mpc^{-3} \, yr^{-1}}\right) = \kappa - \lambda z - \mu e^{z_{0} - z},
\end{equation}
where $\kappa$, $\lambda$, $\mu$, and $z_0$ are parameters, the best-fitting values derived from the usual MCMC procedure for cosmologies are found in Table~\ref{tab:sfr}. Note that the errorbars represent the propagated uncertainties from the previous relations and decrease at lower redshift as we have more statistical power. At $z = 10$, WDM is a factor of two lower than the CDM case, and this factor decreases as redshift decreases. By $z = 7$, the two cosmologies are very similar and essentially indistinguishable at even lower redshift. Both scenarios lie within the range of values between the uncorrected and dust-corrected results from \cite{Bouwens2015} as indicated by the shaded region in the figure. Note that we do not correct for dust or consider any effects beyond our intrinsic star formation.

\begin{table}
\centering
\begin{tabular}{ |c|c|c|c|c| }
\hline
\hline
Cosmology  & $\kappa$ & $\lambda$ & $\mu$ & $z_{0}$ \\
\hline
CDM & 0.00131 & 0.254 & 6.03 & 0.0937 \vspace{0.1cm}\\
WDM & 0.676 & 0.377 & 6.16 & 0.381  \\
\hline
\hline
\end{tabular}
\caption{Best-fitting parameter values for global SFR density, given by equation~(\ref{eq:sfrdens}).}
\label{tab:sfr}
\end{table}

\subsection{Reionization History}
\label{sec:reion}

In this section, we use the predicted galaxy populations in the CDM and WDM models to explore the effects of the WDM cosmology on cosmic reionization, using a commonly used, simple ODE model. We emphasize that this model is a rough estimate of the reionization history, but is a sufficient first step given the uncertainties in reionization modelling. 
In particular, we treat $f_{\rm esc}$, the fraction of ionizing photons that escape from galaxies into the IGM, as a free parameter that is tuned to match current observational measurements.

We follow the approach described in \cite{Kuhlen2012}, starting with this differential equation for the ionized fraction: 

 \begin{equation}
 \label{eq:ionfrac}
 \dot{Q}_{\ion{H}{ii}} = \frac{\dot{n}_{\rm ion}}{\langle n_{\rm H} \rangle} - \frac{Q_{\ion{H}{ii}}}{t_{\rm rec}}.
 \end{equation}
Here, $\langle n_{\rm H} \rangle = X_{\rm p} \Omega_{\rm b} \rho_{\rm c}$ is the comoving hydrogen density, in terms of the hydrogen mass fraction $X_{\rm p} = 0.75$, the baryon density $\Omega_{\rm b} = 0.049$, and the critical density $\rho_{\rm c}$. The IGM recombination time is:

 \begin{align}
  t_{\rm rec} & = \left[C_{\ion{H}{ii}}\alpha_{\rm B}(T)\left(1+\frac{Y_{\rm p}}{4X_{\rm p}}\right)\langle n_{\rm H} \rangle (1+z)^3\right]^{-1} \\
	         &  \approx 0.97\,{\rm Gyr} \left(\frac{C_{\ion{H}{ii}}}{3}\right)^{-1} \left(\frac{T}{2 \times 10^{4}\,{\rm K}} \right)^{0.7}  \left(\frac{1+z}{7} \right)^{-3},
 \end{align}
where $\alpha_{\rm B}(T) $ is the case-B recombination coefficient at an IGM temperature $T = 2 \times 10^4$~K, with the value of $1.6 \times 10^{-13}$~cm$^3$~s$^{-1}$ \citep{Storey1995}, and $Y_{\rm p} = 0.25$ being the helium mass fraction. $C_{\ion{H}{ii}}$ is the effective clumping factor of ionized gas in the IGM; we adopt a constant value of 3 in this work. Of course, a lower value would decrease the resultant $f_{\rm esc}$ and vice versa. For example, $C_{\ion{H}{ii}}$ = 1 requires an $f_{\rm esc}$ that is approximately 10 per cent lower.

Finally, $\dot{n}_{\rm ion}$ is the global production rate of ionizing photons:

\begin{equation}
 \dot{n}_{\rm ion} = f_{\rm esc}\ \xi_{\rm ion}\ \rho_{\rm SFR},
\end{equation}
where $f_{\rm esc}$ is the effective fraction of photons that escape from galaxies to reionize the IGM, and $ \xi_{\rm ion}$  is the ionizing photon production efficiency for a typical stellar population per unit time per unit SFR:

\begin{equation}
\rm log \left(\frac{\xi_{ion}}{photons\ s^{-1}\ M_{\odot}\ yr} \right) = 53.14, 
\end{equation}
where the right-hand value is chosen to be consistent with recent works, e.g. \cite{Lovell2018}. In fact, some observational work points to a higher value \citep[e.g.][]{Topping2015}. The exact value is unimportant as $\xi_{\rm ion}$ is completely degenerate with $f_{\rm esc}$, and a higher $\xi_{\rm ion}$ merely indicates a lower $f_{\rm esc}$ and vice versa.

\begin{figure}
\centering
 	\includegraphics[width=85mm]{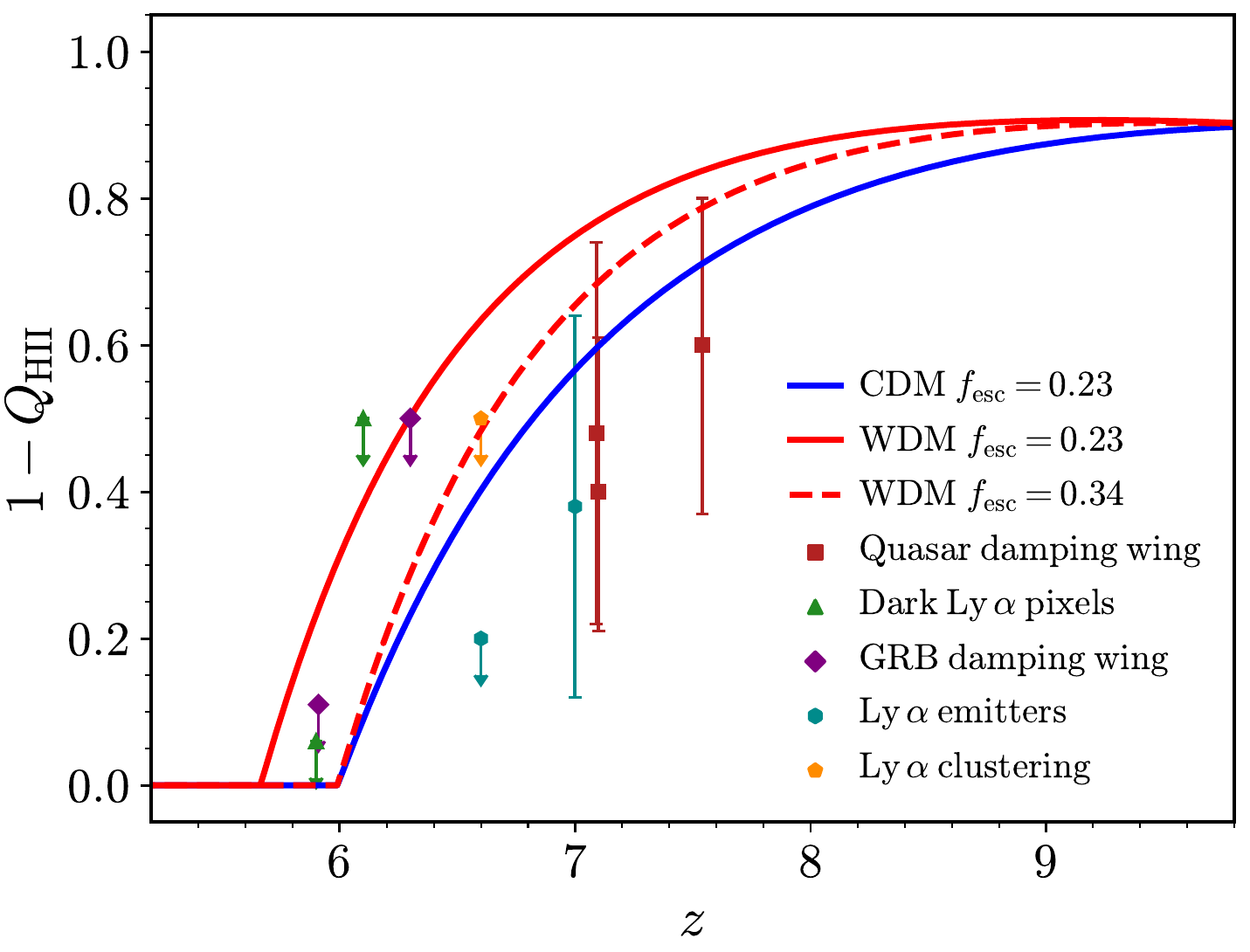}
 	\caption{Neutral fraction of hydrogen as a function of redshift. The blue solid line shows the CDM ionization with a tuned $f_{\rm esc} = 0.23$. The red solid line shows the ionization in WDM with the same escape fraction. The red dashed line shows the WDM ionization forced to match CDM results with $ f_{\rm esc} = 0.34$. Our results are compared to observational inferences from Ly~$\alpha$ damping wings (squares; \protect\citealt{Greig2017,Davies2018,Greig2019}), dark Ly $\rm \alpha$ forest pixels (triangles; \protect\citealt{McGreer2011, McGreer2015}), GRB damping wing absorption (diamonds; \protect\citealt{McQuinn2008, Chornock2013}), decline in Ly $\rm \alpha$ emitters (hexagons; \protect \citealt{Ota2008, Ouchi2010}), and Ly $\rm \alpha$ clustering (pentagons; \protect\citealt{Ouchi2010})}
 	 \label{fig:ionfrac}
 \end{figure}

The differential equation was evolved backwards in time, starting with an ionized fraction of $0.1$ at $z = 10$, a value consistent with results from simulations, e.g. \cite{Dixon2016}, and observations of the CMB optical depth measurements \citep{Planck2016}. This simplification was necessary due to our small sample of sufficiently massive haloes at $z > 10$, limiting the accuracy of predicted stellar mass functions (see Fig.~\ref{fig:hmf}). The $f_{\rm esc}$ was treated as a free parameter and tuned separately for the CDM and WDM scenarios with the goal of matching observations and reaching $Q_{\ion{H}{ii}}(z=6) = 1$. We found that in CDM $f_{\rm esc} = 0.23$ matched the observed data well, while WDM can be forced to match CDM with an increased $f_{\rm esc} = 0.34.$ Fig.~\ref{fig:ionfrac} summarizes our ionization history results. The initial ionization fraction assumption does not significantly impact our results. For example, doubling the initial ionized fraction decreases the resultant $f_{\rm esc}$ at the percent level. 

The physical value of $f_{\rm esc}$, including the potential redshift and halo mass dependence, is a hotly debated topic. Measurements of nearby galaxies have yielded very small values of a few per cent \citep[e.g.][]{Rutkowski2017,Steidel2018}. \cite{Faisst2016} have observed a somewhat larger $f_{\rm esc}$ and evidence of increasing values with redshift. Simulations have not reached a consensus, but $f_{\rm esc}$ appears to vary significantly from halo to halo \citep[e.g.][]{Paardekooper2015,Trebitsch2017,Rosdahl2018}. Although our values are generally larger, the high redshift Universe remains uncertain, and $f_{\rm esc}$ and $\xi_{\rm ion}$ are degenerate, meaning a larger $\xi_{\rm ion}$ could bring our results more in line with observations.

\begin{figure}
 \centering
 	\includegraphics[width=85mm]{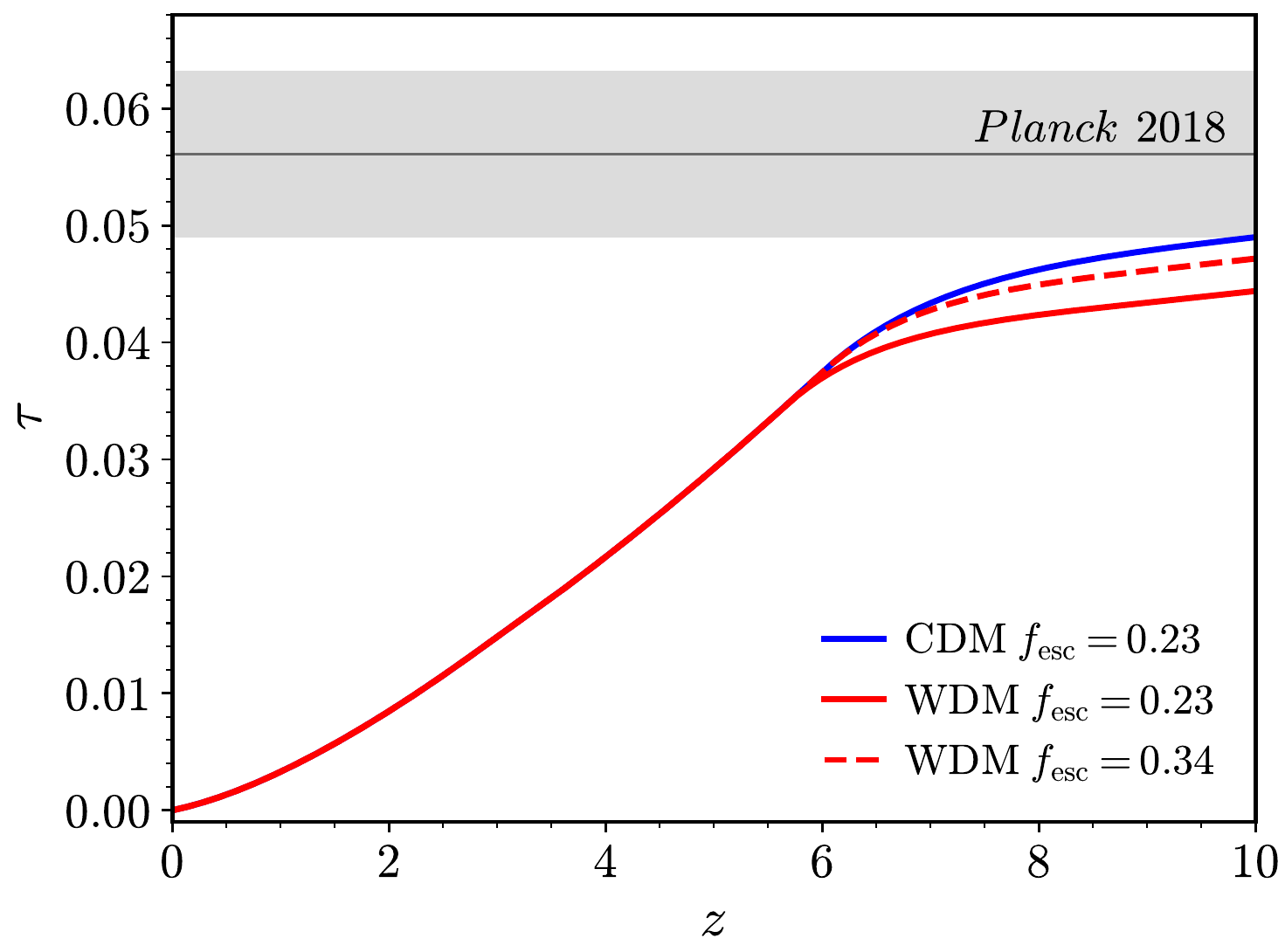}
 	\caption{The CMB optical depth as a function of redshift. The blue solid line shows the CDM ionization with a tuned $f_{\rm esc} = 0.23$. The red solid line shows the ionization in WDM with the same escape fraction. The red dashed line shows the WDM ionization forced to match CDM results with $ f_{\rm esc} = 0.34$. The shaded area shows the 1$\sigma$ confidence interval from \protect \cite{Planck2018} with the thin line representing the best fit from all combined data. }
 	\label{fig:optdepth}
 \end{figure}

Finally, we used the resultant ionized fractions to compute the CMB optical depth:

\begin{equation}
\tau(z) = c \langle n_{\rm H} \rangle \sigma_{\rm T} \int_{0}^{z} f_{\rm e} Q_{\ion{H}{ii}}(z') H^{-1}(z')(1+z')^2 dz',
\end{equation}
where $c$ is the speed of light; $\sigma_{\rm T}$ is the Thompson cross section; $\rm H(z)$ is the Hubble parameter. $f_{\rm e} = 1\ + \ \eta\ Y_{\rm p}/4X_{\rm p}$ is the number of free electrons per hydrogen nucleus. We consider helium to be singly ionized $(\eta = 1)$ at the same rate as hydrogen for $z > 3$ and doubly ionized $(\eta = 2)$ at $z\leq 3$, which is consistent with recent observations \citep{Worseck2018}. Fig.~\ref{fig:optdepth} shows our optical depth results. 

The WDM 3~keV model struggles to reach the $1\sigma$ confidence level of the Planck measurement for optical depth with the CDM-inspired $f_{\rm esc} = 0.23$. This result is a direct consequence of the delay in reionization shown in Fig.~\ref{fig:ionfrac}. The low optical depth even for our CDM results can be partially explained by the lack of simulation data at $z > 10$. Furthermore, reionization modelling that includes inhomogeneity and more physical processes will generically produce slower reionization histories and therefore a larger $\tau$.

\section{Conclusion}
\label{sec:concl}

With upcoming facilities such as ALMA and {\it JWST}, we will have access to unprecedented data for the high-redshift Universe and the formation and properties of the first galaxies.
The data are sensitive to the early phase of structure formation and hence might help shed light on the nature of dark matter.
In this paper, we present a set a large set of high-resolution cosmological hydrodynamical simulations of galaxy formation performed in two different cosmological models, Cold Dark Matter (CDM) 
and Warm Dark Matter (WDM). For the WDM simulations, we have chosen a particle mass of 3~keV, in agreement with current constraints \citep{Irsic2017}.

Our simulations cover the redshift range $5 < z < 10$ and span three orders of magnitude in halo mass, from $10^8$ to $10^{11}$~\Msun (at $z=6$) and more than four in stellar mass, 
for a total of 38 individual zoom-in simulations among the two cosmologies. In addition to the main zoom-in galaxy, we include all haloes with more than 10,000 particles, resulting in over 100 galaxies at $z = 5$.
Simulations in the two different cosmologies have been performed with the same implementation of the baryonic physics as in the NIHAO project \citep{Wang2015}, which has proven to
be extremely successful in reproducing several observed properties of galaxies at low and intermediate redshift  \citep{Wang2015, Dutton2017} and 
for a very extended mass range \cite[e.g.][]{Maccio2017, Buck2018}.

WDM simulations have a systematically lower stellar mass function at low masses. This deficit is due to a combination of three effects:  
a suppression of the number of haloes with mass less than  $\approx 10^9~\Msun$, a slightly
lower star formation efficiency at a fixed halo mass, and a larger fraction of dark haloes for masses below $\approx 10^9~\Msun$.
At all redshifts, the two mass functions start to depart at a stellar mass of $10^7~\Msun$. The difference
peaks at about an order of magnitude at $M_{\star} \approx 10^{5.5}~\Msun$ and is more pronounced at higher redshift.
Unfortunately, for a realistic candidate of 3~keV, these differences are on the mass scales not yet accessible by current facilities, such as \textit{Hubble Space Telescope}. 

The delayed structure formation in WDM models also manifests itself in the global star formation density rate (star formation rate per unit volume) as a function of redshift, which is 
about a factor of a few lower in WDM w.r.t. CDM at high redshift. By $z = 6$, this quantity is very similar in these two models.

Finally, the different star formation histories in the two cosmologies also leave an imprint in the reionization history of the Universe. For a fixed photon escape fraction ($f_{\rm esc}=0.23$), 
the fraction of neutral hydrogen is higher by ~0.3 in WDM for a redshift range between $5 < z < 7$; a larger escape fraction ($f_{\rm esc}=0.34$) is needed in order to reconcile
WDM with current data and have a completely ionized Universe by $z \approx 6$. The same effect is visible in the CMB optical depth, where WDM is inconsistent with Planck constraints 
for $f_{\rm esc} = 0.23$, and a larger value than even 0.34 is needed to match the data.

Our results suggest that high redshift is one of the best places to look for the signature of a possible warm candidate on structure formation, in agreement
with recent studies from \cite{Corasaniti2017}.
Even for a quite warm-ish mass of 3~keV, the delayed structure formation of WDM leaves a distinct imprint on observable quantities, such as 
the stellar mass function, and on the evolution of cosmic reionization. On the other hand, as recently shown by \cite{Villanueva2018}, a firm comparison with current data is at the moment 
prevented by our lack on knowledge on how to parametrize the `baryonic' side of structure formation, as for example the UV photon escape fraction. Our results, combined with previous works, strengthen the need to combine galaxy formation and cosmology if we want to shed light on the nature
of the dark components of our Universe.

 \section*{Acknowledgments} 

The authors  gratefully acknowledge the Gauss Centre for Supercomputing e.V. (www.gauss-centre.eu) 
for funding this project by providing computing time on the GCS Supercomputer SuperMUC at the Leibniz Supercomputing Centre (www.lrz.de) and 
the High Performance Computing resources at New York University Abu Dhabi.
 

\vspace{-0.5cm}
\bibliographystyle{mnras}
\bibliography{bibliography}

\appendix
\newpage
\section{Parameter fitting values}
\label{app:hmf}

In this appendix, we present the parameters derived from the MCMC fitting procedure for the majority of relations presented in the text. Table~\ref{tab:hmf} shows our best-fittingting parameters for the halo mass function at $z = 5-10$ in CDM and 3~keV WDM cosmologies, according to equation~(\ref{eq:hmf}). Table~\ref{tab:moster} includes the median and 1$\sigma$ parameter values for the $M_\star$-$M_{200}$ relation, which in linear in log-log space. Table~\ref{tab:fstar} gives the same quantities for the fraction of dark galaxies, as described by equation~(\ref{eq:fstar}).

\begin{table}
\centering
\begin{tabular}{ |c|c|c|c|c|c|c| }
\hline
\hline
$z$ & Cosmology  & $A$ & $B$ & $C$ & $M_{0}$ & $\alpha$ \\
\hline
\hline
\multirow{2}{.3cm}{5} & CDM & 52.93 & 3.641 & 15.31 & 11.05 &  0.1315 \\
& WDM & 48.99 & 3.863 & 16.06 & 8.574 & 0.2371 \\
\hline
\multirow{2}{.3cm}{6} & CDM & 53.23 & 3.851 & 17.12 & 9.568 & 0.1462 \\
& WDM & 53.60 & 4.331 & 14.72 & 9.092 & 0.2569 \\
\hline
\multirow{2}{.3cm}{7} & CDM & 47.86 & 3.819 & 14.86 & 8.572 & 0.1862 \\
& WDM & 62.80 & 5.0117 & 17.48 & 9.201 & 0.2411 \\
\hline
\multirow{2}{.3cm}{8} & CDM & 54.53 & 4.338 & 16.04 & 8.975 & 0.1837 \\
& WDM & 60.06 & 4.987 & 14.13 & 9.495 & 0.2628 \\
\hline
\multirow{2}{.3cm}{9} & CDM & 51.62 & 4.274 & 15.18 & 8.557 & 0.1963 \\
& WDM & 58.73 & 4.960 & 15.83 & 8.878 & 0.2616 \\
\hline
\multirow{2}{.3cm}{10} & CDM & 54.11 & 4.652 & 11.09 & 9.837 & 0.2199 \\
& WDM & 65.11 & 5.537 & 21.02 & 8.175 & 0.2643 \\
\hline
\hline
\end{tabular}
\caption{Best-fitting parameter values for halo mass functions.}
\label{tab:hmf}
\end{table}

\begin{table}
\centering
\begin{tabular}{ |c|c|c|c| }
\hline
\hline
$z$ & Cosmology  & slope & intercept \\
\hline
\hline
\multirow{2}{.3cm}{5} & CDM & $1.536_{-0.051}^{+0.049}$ & $-8.24_{-0.47}^{+0.50}$ \vspace{0.2cm} \\ 
& WDM & $1.683_{-0.062}^{+0.058}$ & $-9.73_{-0.58}^{+0.61}$  \\
\hline
\multirow{2}{.3cm}{6} & CDM & $1.490_{-0.052}^{+0.050}$ & $-7.67_{-0.47}^{+0.50}$  \vspace{0.2cm} \\
& WDM & $1.621_{-0.070}^{+0.065}$ & $-9.01_{-0.63}^{+0.67}$  \\
\hline
\multirow{2}{.3cm}{7} & CDM & $1.448_{-0.060}^{+0.056}$ & $-7.25_{-0.51}^{+0.56}$  \vspace{0.2cm} \\
& WDM & $1.483_{-0.082}^{+0.078}$ & $-7.61_{-0.74}^{+0.78}$   \\
\hline
\multirow{2}{.3cm}{8} & CDM & $1.358_{-0.069}^{+0.064}$ & $-6.37_{-0.58}^{+0.62}$ \vspace{0.2cm} \\
& WDM & $1.444_{-0.100}^{+0.094}$ & $-7.23_{-0.87}^{+0.93}$  \\
\hline
\multirow{2}{.3cm}{9} & CDM & $1.354_{-0.083}^{+0.078}$ & $-6.25_{-0.96}^{+0.73}$  \vspace{0.2cm} \\
& WDM & $1.350_{-0.13}^{+0.13}$ & $-6.28_{-1.16}^{+1.21}$  \\
\hline
\multirow{2}{.3cm}{10} & CDM & $1.43_{-0.12}^{+0.11}$ & $-6.96_{-0.99}^{+1.06}$ \vspace{0.2cm} \\
& WDM & $1.41_{-0.20}^{+0.19}$ & $-6.89_{-1.71}^{+1.81}$ \\
\hline
\hline
\end{tabular}
\caption{The median parameter values, including the 1$\sigma$ uncertainties, derived from the MCMC sample for the $M_\star$-$M_{200}$ relation in log-log space. }
\label{tab:moster}
\end{table}

\begin{table}
\centering
\begin{tabular}{ |c|c|c|c| }
\hline
\hline
$z$ & Cosmology  & $\beta$ & $M_1$ \\
\hline
\hline
\multirow{2}{.3cm}{5} & CDM & $1.45_{-0.32}^{+0.63}$ & $-8.73_{-0.17}^{+0.16}$ \vspace{0.2cm} \\ 
& WDM & $1.64_{-0.47}^{+1.04}$ & $-8.96_{-0.17}^{+0.19}$  \\
\hline
\multirow{2}{.3cm}{6} & CDM & $1.82_{-0.42}^{+0.88}$ & $-8.46_{-0.12}^{+0.15}$  \vspace{0.2cm} \\
& WDM & $1.82_{-0.50}^{+1.05}$ & $-8.62_{-0.16}^{+0.19}$  \\
\hline
\multirow{2}{.3cm}{7} & CDM & $2.02_{-0.48}^{+0.96}$ & $-8.35_{-0.12}^{+0.14}$  \vspace{0.2cm} \\
& WDM & $2.08_{-0.62}^{+1.22}$ & $-8.59_{-0.16}^{+0.22}$   \\
\hline
\multirow{2}{.3cm}{8} & CDM & $2.43_{-0.63}^{+1.15}$ & $-8.15_{-0.13}^{+0.13}$ \vspace{0.2cm} \\
& WDM & $2.50_{-0.76}^{+1.31}$ & $-8.20_{-0.22}^{+0.27}$  \\
\hline
\multirow{2}{.3cm}{9} & CDM & $2.36_{-0.58}^{+1.11}$ & $-7.99_{-0.14}^{+0.10}$  \vspace{0.2cm} \\
& WDM & $2.74_{-0.88}^{+1.30}$ & $-7.99_{-0.30}^{+0.30}$  \\
\hline
\multirow{2}{.3cm}{10} & CDM & $2.55_{-0.69}^{+1.19}$ & $-8.03_{-0.15}^{+0.12}$ \vspace{0.2cm} \\
& WDM & $2.78_{-0.94}^{+1.29}$ & $-7.97_{-0.38}^{+0.37}$ \\
\hline
\hline
\end{tabular}
\caption{The median parameter values, including the 1$\sigma$ uncertainties, resulting from the MCMC procedure for $f_{\star}$, which follows the functional form of equation~(\ref{eq:fstar}). }
\label{tab:fstar}
\end{table}

\begin{table}
\centering
\begin{tabular}{ |c|c|c| }
\hline
\hline
$z$  & slope & intercept \\
\hline
\hline
5 & 0.999 & $-8.947$ \\
\hline
6 & 1.019 & $-8.493$ \\
\hline
7 & 1.070 & $-8.802$ \\
\hline
8 & 1.030 & $-8.408$ \\
\hline
9 & 0.9930 & $-8.053$ \\
\hline
10 & 0.9994 & $-8.026$ \\
\hline
\hline
\end{tabular}
\caption{The best-fitting parameters derived from the MCMC procedure for the SFR-$M_{\star}$ relation, which is linear in log-log space. }
\label{tab:sfr}
\end{table}

\end{document}